\documentclass[draftclsnofoot,onecolumn, 12pt]{IEEEtran}
\usepackage{graphicx}
\usepackage{epstopdf}
\usepackage{graphicx}
\usepackage{amsmath}
\usepackage{amsfonts}
\usepackage{amssymb, color}
\usepackage{authblk}
\newtheorem{theorem}{Theorem}

\newtheorem{corollary}{Corollary}
\newtheorem{remark}{Remark}
\begin{document}

%-------------------------- paper title--------------------------------------------%
\title{Capacity of a Full-Duplex Wirelessly Powered Communication System with Self-Interference and Processing Cost}
\author{Ivana Nikoloska, \emph{Student Member, IEEE}, Nikola Zlatanov, \emph{Member, IEEE}, \thanks{Ivana Nikoloska and Nikola Zlatanov are with the Department of Electrical and Computer Systems Engineering, Monash University, Melbourne, Australia. Emails: ivana.nikoloska@monash.edu, nikola.zlatanov@monash.edu} and Zoran Hadzi-Velkov, \emph{Senior Member, IEEE} \thanks{Zoran Hadzi-Velkov is with the Faculty of Electrical Engineering and Information Technologies, Ss. Cyril and Methodius University, Skopje, Macedonia. Email: zoranhv@feit.ukim.edu.mk}}
\maketitle

%-------------------------- abstract--------------------------------------------%
\begin{abstract}
In this paper, we investigate the capacity of a point-to-point, full-duplex (FD), wirelessly powered communication system impaired by self-interference. This system is comprised of an energy transmitter (ET) and an energy harvesting user (EHU), both operating in a FD mode. The ET transmits energy towards the EHU. The EHU harvests this energy and uses it to transmit information back to the ET. As a result of the FD mode, both nodes are affected by self-interference. The self-interference has a different effect at the two nodes: it impairs the decoding of the received signal at the ET, however, it provides an additional source of energy for the EHU. This paper derives the capacity of this communication system assuming a processing cost at the EHU and additive white Gaussian noise channel with block fading. Thereby, we show that  the capacity achieving scheme is relatively simple and therefore applicable to devices with limited resources. Moreover, our numerical results show significant improvements in terms of data rate when the capacity achieving strategy is employed compared to half-duplex transmission. Moreover, we show the positive and negative effects of the self-interference at the EHU and the ET, respectively. Furthermore, we show the crippling effect of the processing cost and  demonstrate that failing to take it into consideration gives a false impression in terms of achievable rate.
\end{abstract}

%------------------------ Introduction--------------------------------------------%
\section{Introduction}\label{Sec-Intro}
In recent years, wireless communication, among many others, has been highly affected by emerging green technologies, such as green radio communications and energy harvesting (EH). While the former aims at minimizing the use of precious radio resources, the latter relies on harvesting energy  from renewable and environmentally friendly sources such as, solar, thermal, vibration or wind, [\ref{intro_lit1}], [\ref{intro_lit2}], and using this energy for the transmission of information. Thereby, EH promises a perpetual operation of communication networks. Moreover, EH completely eliminates the need for battery replacement and/or using power cords, making it a highly convenient option for communication networks with nodes which are hard or dangerous to reach. This is not to say that EH communication networks lack their own set of challenges. For example, ambient energy sources are often intermittent and scarce, which can put in danger the continuous reliability of the communication session. Possible solution  to this problem is harvesting of radio frequency (RF) energy from a dedicated   energy transmitter, which gives rise to the so called wireless powered communication networks (WPCNs) [\ref{intro_lit3}], [\ref{intro_lit4}].

EH technology and WPCNs are also quite versatile, and have been studied in many different contexts. Specifically, the capacity of the EH additive white Gaussian noise (AWGN) channel where the transmitter has no battery for energy storage, was studied in [\ref{intro_lit4a}]. Transmitters with finite-size batteries have been considered in [\ref{intro_lit4b}], [\ref{intro_lit4c}]. In [\ref{intro_lit4b}], authors use stationarity and ergodicity to derive a series of computable capacity upper and lower bounds for the general discrete EH channel. Authors in [\ref{intro_lit4c}] investigate the capacity of EH binary symmetric channels with deterministic energy arrival processes. Both small and large battery regimes have been investigated in [\ref{intro_lit4d}], [\ref{intro_lit4f}]. The capacity of a sensor node with an infinite-size battery has been investigated in [\ref{intro_lit4g}]. Infinite-size batteries have also been adopted in [\ref{intro_lit5}], [\ref{intro_lit6}]. In [\ref{intro_lit5}], the authors derive the capacity of the EH AWGN channel without processing cost or storage inefficiencies, while the authors in [\ref{intro_lit6}] take into account the processing cost as well as the energy storage inefficiencies. The authors in [\ref{intro_lit7}] provide two capacity achieving schemes for the AWGN channel with random energy arrivals at the transmitter. The outage capacity of a practical EH circuit model with a primary and secondary energy storage devices has been studied in [\ref{intro_lit8}].  The authors in [\ref{intro_lit9}] derive the minimum transmission outage probability for delay-limited information transfer and the maximum ergodic capacity for delay unlimited information transfer versus the maximum average energy harvested at the receiver. The capacity of the Gaussian multiple access channel (MAC) has been derived in [\ref{intro_lit9a}], [\ref{intro_lit9b}]. Relaying (i.e., cooperative) networks with wireless energy transfer have also been extensively analysed due to their ability to guarantee longer distance communication than classical point-to-point EH links [\ref{intro_lit10}], [\ref{intro_lit10a}]. Information theory has also been paired with queuing theory in order to compute the capacity of EH links [\ref{intro_lit10b}].

Most of the considered EH system models in the literature are assumed to operate in the half-duplex (HD) mode. This is logical since up until recently it was  assumed that  simultaneous reception and transmission on the same frequency, i.e., full-duplex (FD) communication is impossible. However, recent developments have shown this assumption to be false, i.e., have shown that FD communication is possible.  To accomplish FD communication, a radio has to completely cancel or significantly reduce the inevitable self-interference. Otherwise, the self-interference increases the amount of noise and  thereby reduces the achievable rate. Research efforts have made significant progress when it comes to tackling this problem and both active and passive cancelation methods have been proposed. The former, refers to techniques which introduce signal attenuation when the signal propagates from the transmit antenna to the receiver one [\ref{intro_lit17a}], [\ref{intro_lit17b}]. The latter exploits the knowledge of the transmit symbols by the FD node in order to cancel the self-interference [\ref{intro_lit17c}]-[\ref{intro_lit17e}]. Combinations of both methods have also been considered [\ref{intro_lit17f}].

Motivated by the idea that the role of the self-interference can be redefined in WPCNs, such as in [\ref{sys_mod_lit1}]-[\ref{intro_lit18c}], as well as by the advances in FD communication, in this work, we investigate the capacity of a FD wirelessly powered communication system comprised of an energy transmitter (ET) and an energy harvesting user (EHU) that operate in an AWGN block-fading environment. In this system, the ET  sends RF energy to the EHU, whereas, the EHU harvests this energy and uses it to transmit information back to the ET. Both the ET and the EHU work in the FD mode, hence, both nodes transmit and receive RF signals in the same frequency band and at the same time. As a result, both are affected by self-interference. The  self-interference has opposite effects at the ET and the EHU. Specifically, the self-interference signal has a negative effect at the ET since it hinders the decoding of the information signal received from the EHU. However, at the EHU, the self-interference signal has a positive effect since it increases the amount of energy that can be harvested by the EHU. In this paper, we derive the capacity of this system model. Thereby, we prove that the capacity achieving distribution at the EHU is Gaussian, whereas, the input distribution at the ET is degenerate and takes only one value with probability one. This allows simple capacity achieving schemes to be developed, which are applicable to nodes with limited resources such as EH devices. Our numerical results show significant gains in term of data rate when the capacity achieving scheme is employed, as opposed to a HD benchmark scheme.

The results and schemes presented in this paper provide clear guidelines for increasing the data rates of future wirelessly powered communication systems. Moreover, due to their relative simplicity, the proposed schemes are applicable to future Internet of Things systems.

The rest of the paper is organized as follows. Section~II provides the system and channel models as well as the energy harvesting policy. Section~III presents the capacity as well as the optimal input probability distributions at the EHU and the ET. Section~IV provides the converse as well as the achievability of the capacity. In Section~V, we provide numerical results and a short conclusion concludes the paper in Section~VI. Proofs of theorems are provided in the Appendix.
%--------------- paragraph #1

%-------------System Model And Problem Formulation------------------------------------%
\section{System Model and Problem Formulation}\label{Sec-Sys}

We consider a system model comprised of an EHU  and an ET, c.f. Fig. 1.  The ET transmits RF energy to the EHU and simultaneously receives information from the EHU. On the other hand, the EHU   harvests the energy  transmitted from the ET and uses it to transmit information back to the ET.

\begin{figure}[tbp]
\centering
\includegraphics[width=3.7in]{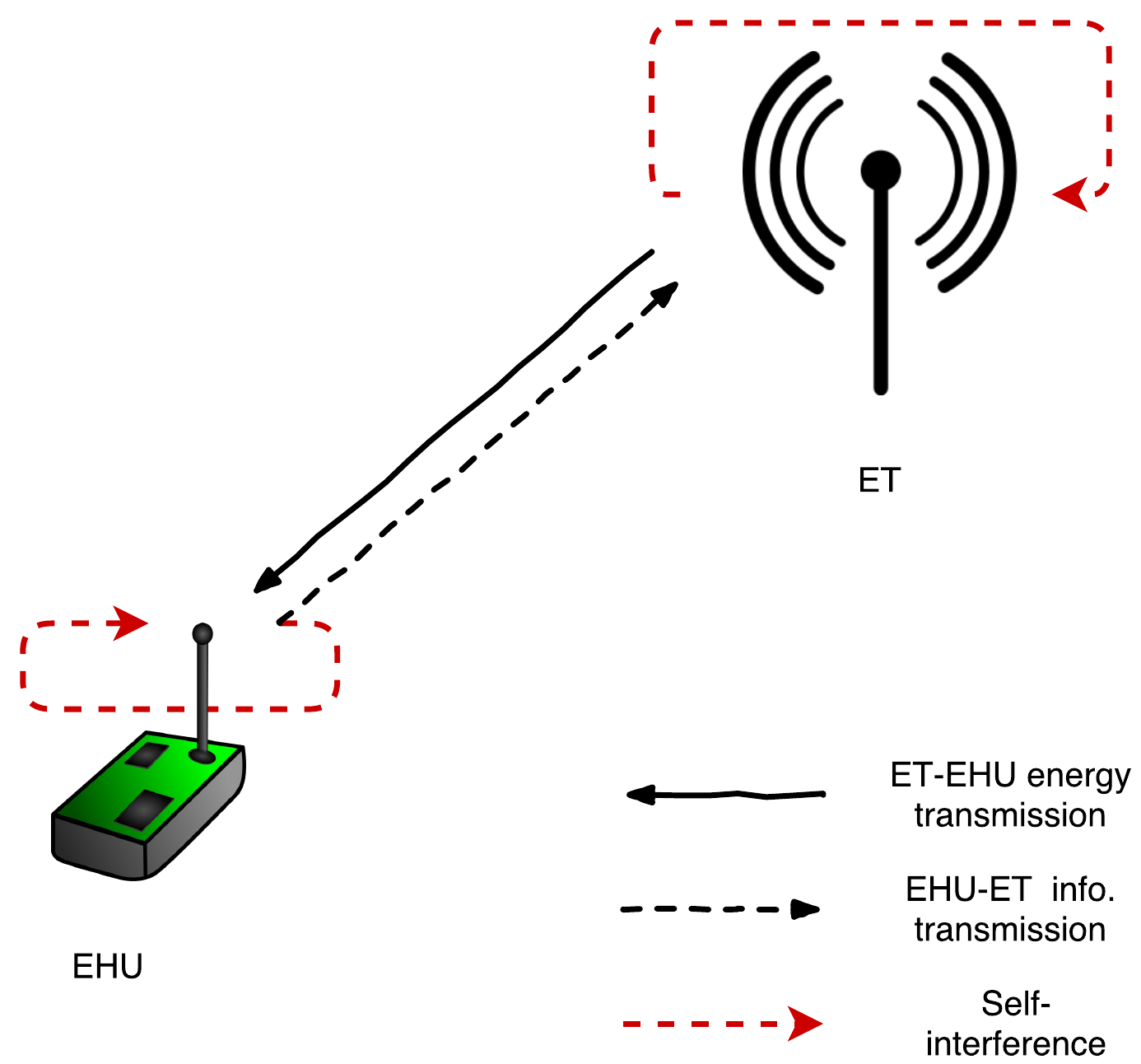}
\caption{System model.}
\label{fig1}
\end{figure}

In order to improve the spectral efficiency of the considered  system, both the EHU and the ET are assumed to operate in the FD mode, i.e., both nodes transmit and receive RF signals simultaneously and in the same frequency band. Thereby, the EHU receives energy signals from the ET and simultaneously transmits information signals to the ET  in the same frequency band. Similarly, the ET transmits energy signals to the EHU and simultaneously receives information signals from the EHU  in the same frequency band. Due to the FD mode of operation, both the EHU and the ET are impaired by self-interference. The self-interference has opposite effects at the ET and the EHU. More precisely, the self-interference signal has a negative effect at the ET since it hinders the decoding of the information signal received from the EHU. As a result, the ET should be designed with a self-interference suppression apparatus, which can suppress the self-interference at the ET and thereby  improve the decoding of the desired signal received from the EHU.  On the other hand, at the EHU, the self-interference signal has a positive effect since it increases the amount of energy that can be harvested by the EHU.  Hence, the EHU should be designed without  a self-interference suppression apparatus  in order for the energy contained in the self-interference signal of the EHU to be fully harvested by the EHU, i.e., the EHU should perform  energy recycling  as proposed in [\ref{sys_mod_lit1}].

\subsection{Channel Model}

We assume that both the EHU and the ET are impaired by AWGNs, with variances $\sigma_1^2$ and $\sigma_2^2$, respectively. Let $H_{12i}$ and $H_{21i}$ model the fading  channel gains of the  EHU-ET and  ET-EHU channels in channel use $i$, respectively. We assume that the channel gains, $H_{12i}$ and $H_{21i}$, follow a block-fading model, i.e., they remain constant during all channel uses in one block, but change from one block to the next, where each block consists of (infinitely) many channel uses. Due to the FD mode of operation, the channel gains $H_{12i}$ and $H_{21i}$ are assumed to be identical, i.e., $H_{12i}$ $=$ $H_{21i}$ $=$ $H_{i}$. We assume that the EHU and the ET are able to estimate the channel gain $H_{i}$ perfectly.

In the $i$-th channel use, let the transmit symbols at the EHU and the ET be modeled as random variables (RVs), denoted by $X_{1i}$ and $X_{2i}$, respectively. Moreover, in channel use $i$, let the received symbols at the EHU and the ET be modeled as RVs, denoted by $Y_{1i}$ and $Y_{2i}$, respectively. Furthermore, in channel use $i$, let the RVs modeling the AWGNs at the   EHU and the ET be denoted by $N_{1i}$ and $N_{2i}$, respectively, and let the RVs modeling the additive self-interferences at the EHU and the ET by denoted by $I_{1i}$ and $I_{2i}$, respectively. As a result,  $Y_{1i}$ and $Y_{2i}$ can be written as
\begin{align}
Y_{1i}&=H_{i} X_{2i}+I_{1i}+N_{1i},\label{eq_1i}\\
Y_{2i}&=H_{i} X_{1i}+I_{2i}+N_{2i}.\label{eq_2i}
\end{align}
We adopt the first-order approximation for the self-interference, justified in [\ref{lit18}] and [\ref{lit19a}], where the authors show that the higher-order terms carry significantly less energy than the first-order terms. Thereby, we only consider the linear components of the self-interference and, similar to [\ref{lit19a}], approximate $I_{1i}$ and $I_{2i}$ as
\begin{align}
&I_{1i}=\tilde G_{1i} X_{1i},\label{eq_si_1a}\\
&I_{2i}=\tilde G_{2i} X_{2i},\label{eq_si_2a}
\end{align}
where $\tilde G_{1i}$ and $\tilde G_{2i}$ denote the self-interference channel gains in channel use $i$. The self-interference channel gains, $\tilde G_{1i}$ and $\tilde G_{2i}$, are time-varying and the statistical properties of these variations are dependent of the hardware configuration and the adopted self-interference suppression scheme. In order for us to obtain the worst-case performance in terms of self-interference, we assume that  $\tilde G_{1i}$ and $\tilde G_{2i}$ are independent and identically distributed (i.i.d.) Gaussian RVs across different channels uses, which is a direct consequence of the fact that the Gaussian distribution has the largest entropy under a second moment constraint, see [\ref{lit19}]. Hence, we assume that $\tilde G_{1i}\sim\mathcal{N}\{\bar g_1,\alpha_1\}$ and $\tilde G_{2i}\sim\mathcal{N}\{\bar g_2,\alpha_2\}$, where $\mathcal{N}\{\bar g,\alpha\}$ denotes a Gaussian distribution with mean $\bar g$ and variance $\alpha$.

 Inserting (\ref{eq_si_1a}) and (\ref{eq_si_2a}) into (\ref{eq_1i}) and (\ref{eq_2i}), respectively, we obtain
\begin{align}
Y_{1i}&=H_{i} X_{2i}+\tilde G_{1i} X_{1i}+N_{1i},\label{eq_1a}\\
Y_{2i}&=H_{i} X_{1i}+\tilde G_{2i} X_{2i}+N_{2i}. \label{eq_2a}
\end{align}
Now, since $\tilde G_{1i}$ and $\tilde G_{2i}$ can be written equivalently as $\tilde G_{1i} = G_{1i} + \bar g_1$ and $\tilde G_{2i} = G_{2i} + \bar g_2$, respectively, where $G_{1i}\sim\mathcal{N}\{0,\alpha_1\}$ and $ G_{2i}\sim\mathcal{N}\{0,\alpha_2\}$, without loss of generality, (\ref{eq_1a}) and (\ref{eq_2a}) can also be written equivalently as
\begin{align} \label{eq_3i}
Y_{1i}&=H_{i} X_{2i}+ \bar g_1 X_{1i}+ G_{1i} X_{1i}+N_{1i}
\end{align}
and
\begin{align} \label{eq_4i}
Y_{2i}&=H_{i} X_{1i}+ \bar g_2 X_{2i}+ G_{2i} X_{2i}+N_{2i},
\end{align}
respectively.

Since the ET knows $X_{2i}$ in channel use $i$, and since given sufficient time it can always estimate the mean of its self-interference channel, $\bar g_2$, the ET can remove $\bar g_2 X_{2i}$ from its received symbol $Y_{2i}$, given by (\ref{eq_4i}), and thereby reduce its self-interference. In this way, the ET obtains a new received symbol, denoted again by $Y_{2i}$, as
\begin{align} \label{eq_5i}
Y_{2i}&=H_{i} X_{1i}+   G_{2i} X_{2i}+N_{2i}.
\end{align}
Note that since $G_{2i}$ in (\ref{eq_5i}) changes i.i.d. randomly from one channel use to the next, the ET cannot estimate and remove  $G_{2i} X_{2i}$ from its received symbol. Thus, $G_{2i} X_{2i}$ in (\ref{eq_5i}) is the residual self-interference at the ET. On the other hand, since the EHU benefits from the self-interference, it does not remove  $\bar g_1 X_{1i}$ from its received symbol  $Y_{1i}$, given by (\ref{eq_3i}), in order to have a self-interference signal with a much higher energy, which it can then harvest. Hence, the received symbol at the EHU is given by (\ref{eq_3i}).

In this paper, we investigate the capacity of a channel given by the input-output relations in (\ref{eq_3i}) and (\ref{eq_5i}), where we are only interested in the  mutual-information between $X_{1}^n$ and $Y_{2}^n$ subject to  an average power constraint on $X_2^n$, where the notation $a^n$ is used to denote the vector $a^n=(a_1,\; a_2,\;...,a_n)$. An equivalent block diagram of the considered system model is presented in Fig~2.
\vspace{3mm}
\begin{figure}[tbp]
\centering
\includegraphics[width=6.7in,scale=7]{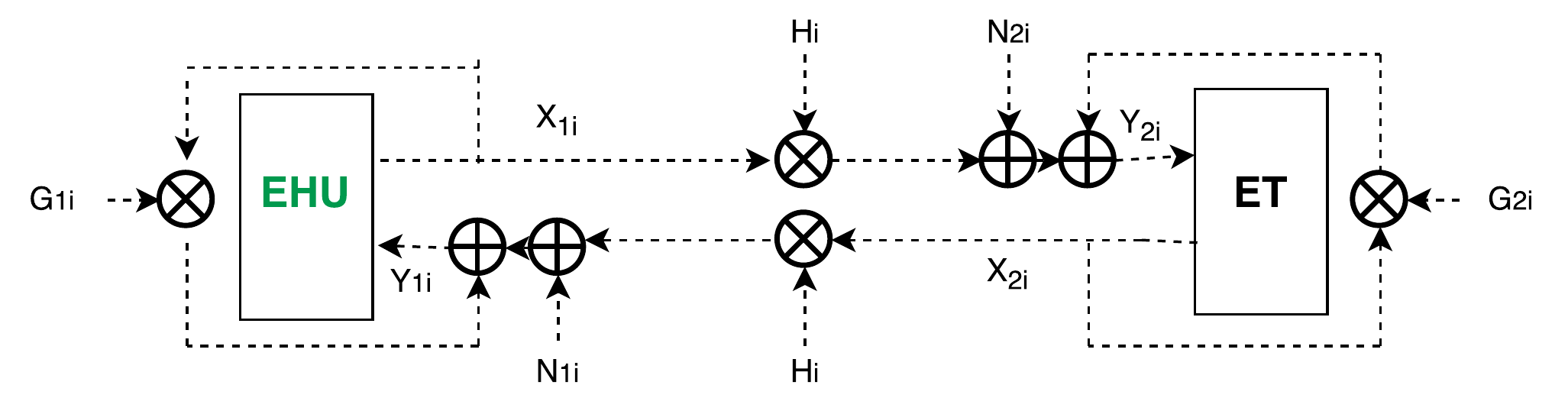}
\caption{Block diagram of the system model.}
\label{fig2}
\end{figure}

\subsection{Energy Harvesting Model}
We assume that the energy harvested by the EHU in channel use $i$ is given by   [\ref{sys_mod_lit1}]
\begin{align}\label{eq_3energy}
%E_i=\text{max} \{ 0, \eta (H_{i}X_{2i} + \bar g_{1} X_{1i}+G_{1i} X_{1i})^2-P_p \},
E_{{\rm in,}i}= \eta (H_{i}X_{2i} + \bar g_{1} X_{1i}+G_{1i} X_{1i})^2,
\end{align}
where $0<\eta<1$ is the energy harvesting inefficiency coefficient. The EHU stores $E_{{\rm in,}i}$ in its battery, which is assumed to be infinitely large. Let $B_i$ denote the amount of harvested energy in the battery of the EHU in the $i$-th channel use. Moreover, let $E_{{\rm out,}i}$ be the extracted energy from the battery in the $i$-th channel use. Then, $B_i$, can be written as
\begin{align}\label{eq_4}
B_i=B_{i-1}+E_{{\rm in,}i}-E_{{\rm out,}i}.
\end{align}
Since in channel use $i$ the EHU cannot extract more energy than the amount of energy stored in its battery during channel use $i-1$, the extracted energy from the battery in channel use $i$,  $E_{{\rm out,}i}$, can be obtained as
\begin{align}\label{eq_4a}
E_{{\rm out,}i}=\min\{B_{i-1},X_{1i}^2 + P_p\},
\end{align}
  where   $X_{1i}^2$ is the  transmit energy of the desired transmit symbol in channel use $i$, $X_{1i}$,  and $P_p$ is the processing cost of the EHU. The processing cost, $P_p$, is the energy spent for processing and the energy spent due to the inefficiency and the power consumption of the electrical components in the electrical circuit such as AC/DC convertors and RF amplifiers.

\begin{remark}
Note that the ET also requires energy for processing. However, the ET is assumed to be equipped with a conventional power source which is always capable of providing the processing energy  without interfering with the energy required for transmission.
\end{remark}

We now use the results in [\ref{intro_lit5}] and [\ref{lit19}], where it was proven that if the total number of channel uses satisfies $n \rightarrow \infty$, the battery of the EHU has an unlimited storage capacity, and
\begin{align}\label{eq_4b}
\mathcal{E}\{E_{{\rm in,}i}\}\geq \mathcal{E}\{X_{1i}^2\}+P_p
\end{align}
 holds, where $\mathcal{E}\{\cdot\}$ denotes statistical expectation,  then the number of channel uses in which the extracted energy from the battery is insufficient  and thereby  $E_{{\rm out,}i}=B_{i-1}$ holds is negligible compared to the number of channel uses when the extracted energy from the battery is sufficient and thereby $E_{{\rm out,}i}=X_{1i}^2+P_p$ holds. In other words, when the above three conditions hold, in almost all channel uses there will be enough energy to be extracted from the EHU's battery for both processing, $P_p$, and for the transmission of the desired transmit symbol $X_{1i}$, $X_{1i}^2$.

\section{Capacity}
The channel in Fig. 2, modeled by (\ref{eq_3i}) and (\ref{eq_5i}), is a discrete-time channel with inputs $X_1$ and $X_2$, and outputs $Y_2$ and $Y_1$. Furthermore, the probability of observing $Y_2$ or $Y_1$ is only dependent on the input $X_{1}$ and/or $X_{2}$ at channel use $i$ and is conditionally independent of previous channel inputs and outputs, making the considered channel a discrete-time memoryless channel [\ref{lit20}]. For this channel, we propose the following theorem.

\begin{theorem}
Assuming that the average power constraint at the ET is $P_{ET}$, the capacity of the considered EHU-ET communication channel is given by
\begin{align}\notag
& C=\underset{p(x_1|x_2,h),p(x_2|h)}{\text{max}} \sum_{x_2 \in \mathcal{X}_2} \sum_{h \in \mathcal{H}} I(X_1;Y_2 | X_2=x_2, H=h)p(x_2|h)p(h) \\ \notag
&{\rm{Subject\;\;  to}}  \nonumber\\
&\qquad\qquad{\rm C1:} \sum_{x_2 \in \mathcal{X}_2} \sum_{h \in \mathcal{H}} x_2^2 p(x_2|h)p(h) \leq P_{ET}  \nonumber\\
&\qquad\qquad{\rm C2:} \int_{x_1} \sum_{x_2 \in \mathcal{X}_2} \sum_{h \in \mathcal{H}} (x_1^2 + P_p) p(x_1|x_2,h)p(x_2|h)p(h)dx_1 \leq  \nonumber\\
&\qquad\qquad\qquad \int_{x_1} \sum_{x_2 \in \mathcal{X}_2} \sum_{h \in \mathcal{H}} E_{\rm in} p(x_1|x_2,h)p(x_2|h)p(h)dx_1  \nonumber\\
&\qquad\qquad{\rm C3:} \sum_{x_2 \in \mathcal{X}_2}{p(x_2|h)} = 1  \nonumber\\
&\qquad\qquad{\rm C4:} \int_{x_1}{p(x_1|x_2,h) d x_1} = 1,    \label{cap_eq1}
\end{align}
where $I(;|)$ denotes the conditional mutual information between $X_1$ and $Y_2$ when the EHU knows the transmit symbols of the ET and both the EHU and the ET have full channel state information (CSI). In (\ref{cap_eq1}), lower case letters $x_2$ and $h$ represent realizations of the random variables $X_2$ and $H$ and their support sets are denoted by $\mathcal{X}_2$ and $\mathcal{H}$, respectively. Constraint C1 in  (\ref{cap_eq1}) constrains the average transmit power of the ET to $P_{ET}$, and C2 is due to (\ref{eq_4b}), i.e., due to the fact that EHU has to have harvested enough energy for both processing and transmission of symbol $X_1$. The maximum in the objective function is taken over all possible conditional probability distributions of $x_1$ and $x_2$, given by $p(x_1|x_2,h)$ and $p(x_2|h)$, respectively.
\end{theorem}
\begin{IEEEproof}
The proof is in two parts. In Subsection~\ref{con} we prove the converse and in Subsection~\ref{Ach} we provide the achievability of the capacity given by (\ref{cap_eq1}).
\end{IEEEproof}

\subsection{Optimal Input Distribution and Simplified Capacity Expression}
The optimal input distributions at the EHU and the ET which achieve the capacity in (\ref{cap_eq1}) and the resulting capacity expressions are provided by the following theorem.
\begin{theorem}
We have two cases for the channel capacity and the optimal input distributions.
\\
Case 1: If
\begin{align}\label{cap_eq2a}
&\sum_{h \in \mathcal{H}} \frac{1}{2} \log \left(1+\frac{h^2 P_{EHU}\left(\sqrt{P_{ET}},h\right)}{\sigma_2^2 + P_{ET} \alpha_2}\right)p(h) = \lambda_1 P_{ET}  + \mu_1 \\ \notag
&+\lambda_2  \left(\left(1-\eta(\bar g_1^2 +\alpha_1) \right)\sum_{h \in \mathcal{H}}P_{EHU} \left(\sqrt{P_{ET}},h\right)p(h)-\eta P_{ET}\sum_{h \in \mathcal{H}} h^2 p(h) \right)
\end{align}
holds,
where $\lambda_1$, $\lambda_2$, and $\mu_1$ are the Lagrangian multipliers associated with constraints C1, C2, and C3 in (\ref{cap_eq1}), respectively, the optimal input distribution at the EHU is zero-mean Gaussian with variance $P_{EHU}\left(\sqrt{P_{ET}},h\right)$, i.e., $p(x_1|x_2,h) \sim\mathcal{N} \left(0, P_{EHU}\left(\sqrt{P_{ET}},h\right)\right)$, where
\begin{align}\label{cap_eq1aa}
P_{EHU}\left(\sqrt{P_{ET}},h\right)=\left[ \frac{1}{\lambda_2 (1-\eta(\bar g_1^2 +\alpha_1))}-\frac{\sigma_2^2+P_{ET} \alpha_2}{h^2} \right]^+,
\end{align}
and $\lambda_2$ and is chosen such that
\begin{align} \label{cap_lambda}
 \left(1-\eta(\bar g_1^2 +\alpha_1)\right)\sum_{h \in \mathcal{H}} P_{EHU} \left(\sqrt{P_{ET}},h\right) p(h)+P_p =
 \eta P_{ET} \sum_{h \in \mathcal{H}}  h^2 p(h)
\end{align}
holds.

On the other hand, the optimal input distribution at the ET is given by
\begin{align}\label{cap_eq22a}
p(x_2|h)=\delta \left(x_2-\sqrt {P_{ET}}\right).
\end{align}
Finally, the capacity for this case is given by
\begin{align}\label{cap_eq3}
C=\sum_{h \in \mathcal{H}}\frac {1}{2} \log \left(1+\frac{h^2 P_{EHU}\left(\sqrt{P_{ET}},h\right)}{\sigma_2^2 + P_{ET} \alpha_2}\right)p(h).
\end{align}
\\
Case 2: If (\ref{cap_eq2a}) does not hold, the optimal input distribution at the EHU is zero-mean Gaussian with variance $P_{EHU}(x_0(h),h)$, i.e, $p(x_1|x_2,h) \sim\mathcal{N} \left(0, P_{EHU}\left(x_0(h),h\right)\right)$, where
\begin{align}\label{cap_eq1a}
P_{EHU}(x_0(h),h)=\left[ \frac{1}{\lambda_2 (1-\eta(\bar g_1^2 +\alpha_1))}-\frac{\sigma_2^2+x_0^2(h) \alpha_2}{h^2} \right]^+ ,
\end{align}
where $x_0(h)$ is given by

%\resizebox{0.0001\hsize}{!}\{
\begin{align}\label{cap_eq2b}
%x_0(h)=\sqrt{ \left[ \frac{h^2}{\bar \lambda \alpha_2 e^{W \left(\frac{2 \ln 2 \left((\lambda'-\lambda \eta h^2)\frac{h^2}{\bar \lambda \alpha_2} -1 \right)}{e^{2 \ln 2 \left(1+(\lambda \eta h^2 - \lambda')\frac{\sigma^2}{\alpha_2} + \mu'\right)}}\right)+e^{2 \ln 2 \left(1+(\lambda \eta h^2 - \lambda')\frac{\sigma^2}{\alpha_2}+ \mu'\right)} - \frac{\sigma_2^2 2 \ln 2}{\alpha_2}}}-\frac{\sigma_2^2}{\alpha_2}  \right]^+ },
x_0(h)=\sqrt{ \left[ \frac{h^2 W \left(\frac{2 \ln 2 \left((\lambda_1-\lambda_2 \eta h^2)\frac{h^2}{\lambda_2 (1-\eta(\bar g_1^2 +\alpha_1)) \alpha_2} -1 \right)}{e^{2 \ln 2 \left(1+(\lambda_2 \eta h^2 - \lambda_1)\frac{\sigma^2}{\alpha_2} + \mu_1 \right)}}\right)}{2 \ln 2 ((\lambda_1 - \lambda_2 \eta h^2) h^2 -\lambda_2 (1-\eta(\bar g_1^2 +\alpha_1)) \alpha_2)}-\frac{\sigma_2^2}{\alpha_2}  \right]^+ },
\end{align}
 % \}
and $\lambda_2$ is chosen such that
\begin{align} \label{cap_lambda}
 \left(1-\eta(\bar g_1^2 +\alpha_1)\right) \sum_{h \in \mathcal{H}} P_{EHU} (x_0(h),h) p(h)+P_p =
 \eta \sum_{h \in \mathcal{H}} x_0^2(h)  h^2 p(h)
\end{align}
holds.
In (\ref{cap_eq2b}), $W(.)$ denotes the Lambert W function.

On the other hand, the optimal input distribution at the ET is given by
\begin{align}\label{cap_eq22b}
p(x_2|h)=\delta (x_2-x_0(h)).
\end{align}

Finally, the channel capacity for this case is given by
\begin{align}\label{cap_eq4}
C=\sum_{h \in \mathcal{H}}\frac {1}{2} \log \left(1+\frac{h^2 P_{EHU}\left(x_0(h),h\right)}{\sigma_2^2 + x_0^2(h) \alpha_2}\right)p(h).
\end{align}
\end{theorem}

\begin{IEEEproof}
Please refer to Appendix A.
\end{IEEEproof}

\begin{figure}[tbp]
\centering
\includegraphics[width=5.7in,scale=4]{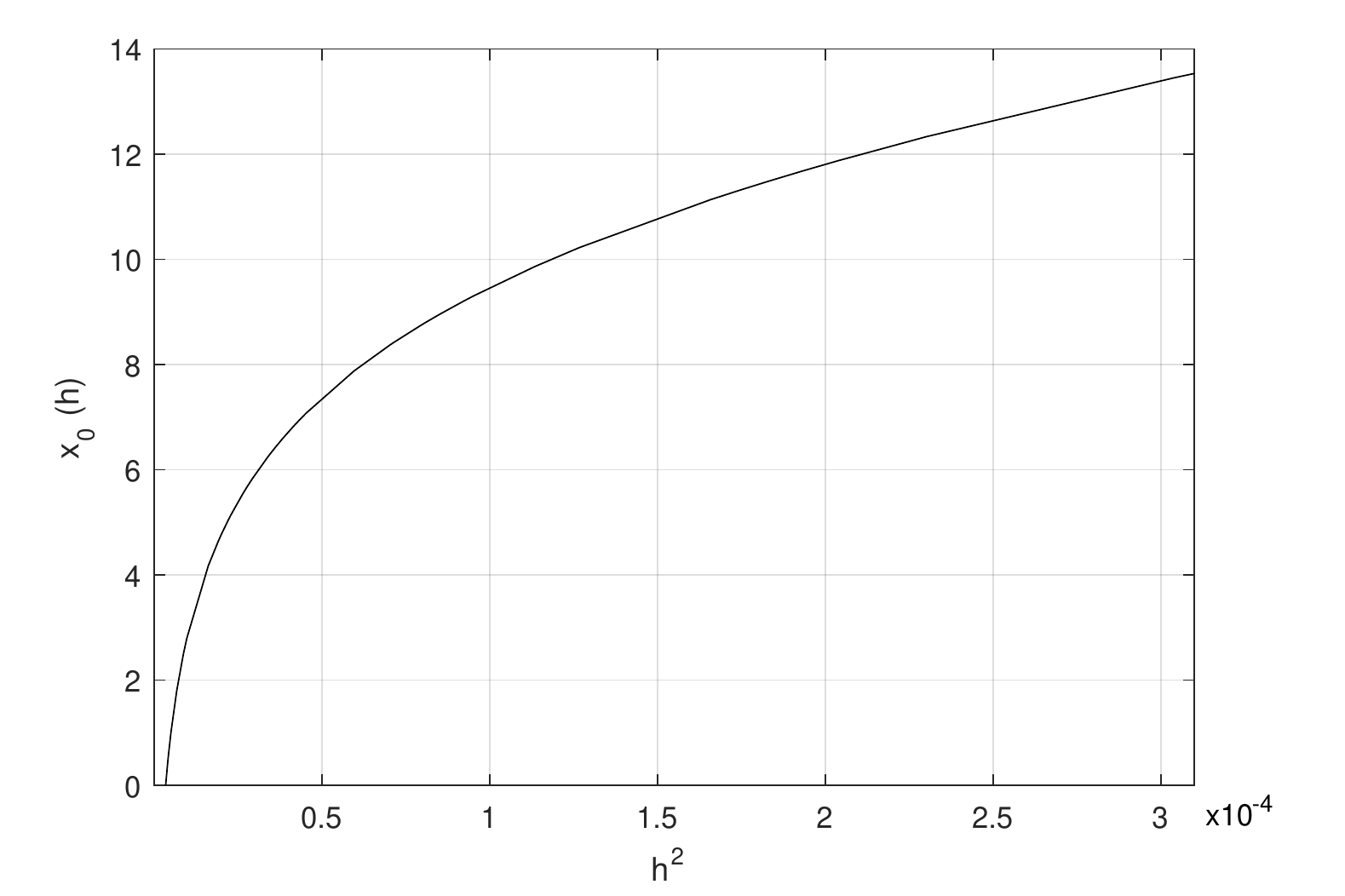}
\caption{ET's transmit symbol $x_0(h)$, given by (\ref{cap_eq2b}), as a function of the fading realisation $h$.}
\label{fig_cap}
\end{figure}

%As can be seen from Theorem 2, the average power of the EHU's transmit symbol, $x_1$, is a function of the ET's transmit symbol, $x_2$, and the fading coefficient, $h$.
Essentially, depending on the average fading power, we have two cases. Case 1 is when the ET transmits the symbol $\sqrt{P_{ET}}$ in each channel use and in all fading blocks. Whereas, the EHU transmits a Gaussian codeword and adapts its power to the fading realisation in each fading block, i.e., it performs classical waterfiling. Thereby,  the stronger the fading channel, $h$, the stronger the average transmit power of the EHU during that fading realisation. Conversely, the weaker the fading channel, $h$,  the lower the average transmit power of the EHU during that fading realisation. In cases when the fading channel is too weak, the EHU remains silent during that fading realisation and only harvests the energy transmitted by the ET. In Case 2, the ET transmits the same symbol $x_0(h)$ during all channel uses of the fading block with fading realisation $h$. Since now $x_0(h)$ depends on the fading, as shown in Fig.~\ref{fig_cap}, the ET transmits different symbols in fading blocks with different fading realisations. In particular, if there is a strong fading gain $h$, $x_0(h)$ is higher. If $h$ is low so is $x_0(h)$. When $h$ is very low, then $x_0(h)=0$, i.e., the ET becomes silent during that fading block. On the other hand, the EHU in this case transmits a Gaussian codeword with an average power $P_{EHU}(x_0(h),h)$, where $P_{EHU}(x_0(h),h)$ is adapted to the fading gain in that fading block. If the fading gain is strong, $P_{EHU}(x_0(h),h)$ is higher and if the fading gain is weak, $P_{EHU}(x_0(h),h)$ is low. If the fading gain is too low, $P_{EHU}(x_0(h),h)=0$, i.e., the EHU becomes silent.

Also,   since $\lambda_2$ is chosen such that constraint C2 in (\ref{cap_eq1}) holds, the expressions in (\ref{cap_eq1aa}), (\ref{cap_eq1a}), and (\ref{cap_eq2b}) are dependent on the processing cost $P_p$. As a result, the optimal solution takes into account $P_p$, which hinders the system's performance, i.e., a larger $P_p$ results in a lower capacity. This is particularly important in networks constrained by their power supply, since their performance in terms of achievable rates and/or lifetime can be easily overestimated by ignoring the processing cost, as shown in the numerical examples, cf. Fig.~\ref{num5}.

\subsection{Special Case of Rayleigh Fading}
When the channel fading gain follows a Rayleigh probability distribution, i.e., their probability density function is given by
\begin{align}\label{spec_eq1}
p_H(h)=\frac{h}{2 \Omega_h} e^{-\frac{h^2}{\Omega_h}},
\end{align}
 the channel capacity in (\ref{cap_eq3}) has a closed-form solution as
\begin{align}\label{spec_eq3}
C=\frac{1}{\ln 2} \left(E_1 \left(\frac{\lambda_2 (1-\eta(\bar g_1^2 +\alpha_1))(\sigma_2^2 + P_{ET} \alpha_2)}{\Omega_h}\right)+e^{-\frac{\lambda_2 (1-\eta(\bar g_1^2 +\alpha_1))(\sigma_2^2 + P_{ET} \alpha_2)}{\Omega_h}} \ln \left(\frac{1}{\Omega_h}\right) \right),
\end{align}
where $\lambda_2$ can be found from the following equation
\begin{align}\label{spec_eq4}
&\frac{1}{\lambda_2 (1-\eta(\bar g_1^2 +\alpha_1))} e^{-\frac{\lambda_2 (1-\eta(\bar g_1^2 +\alpha_1))(\sigma_2^2 + P_{ET} \alpha_2)}{\Omega_h}}-
(\sigma_2^2 + P_{ET} \alpha_2)E_1 \left(\frac{\lambda_2 (1-\eta(\bar g_1^2 +\alpha_1))(\sigma_2^2 + P_{ET} \alpha_2)}{\Omega_h}\right) \nonumber\\
& =\frac{\eta  P_{ET} \Omega_h -P_p}{\left(1-\eta(\bar g_1^2 +\alpha_1)\right) },
\end{align}
where $E_1$ denotes the exponential integral function given by $E_1(x)=\int_{x}^{\infty} \frac{e^{-t}}{t} dt$. When condition (\ref{cap_eq2a}) does not hold, the integral which gives the channel capacity can be evaluated numerically using software such as Mathematica.

\begin{corollary}
When the channel is not impaired by fading, the capacity can be derived as follows

\begin{align}\label{cor_eq1}
C=\frac {1}{2} \log \left(1+\frac{P_{EHU}(\sqrt{P_{ET}})}{\sigma_2^2 + P_{ET} \alpha_2}\right),
\end{align}
where $P_{EHU}(\sqrt{P_{ET}})$ is given by
\begin{align}\label{cor_eq2}
P_{EHU}(\sqrt{P_{ET}})=\frac{\eta P_{ET}}{1-\eta(\bar g_1^2 +\alpha_1)}.
\end{align}
The capacity in (\ref{cor_eq1}) is achieved when $X_1$ follows a Gaussian probability distribution as \\
$p(x_1|x_2) \sim\mathcal{N} \left(0, P_{EHU}\left(\sqrt{P_{ET}}\right)\right)$ and the probability distribution of $X_2$ is degenerate and given by $p(x_2)=\delta \left(x_2-\sqrt {P_{ET}}\right)$.
\end{corollary}

\section{Converse and Achievability of the Channel Capacity}

In this section, we prove the converse and the achievability of the channel capacity.

\subsection{Converse of the Channel Capacity}\label{con}

Let $W$ be the message that the EHU wants to transmit to the ET. Let this message be uniformly selected at random from the message set $\{1,\,2,\,...,2^{nR}\}$, where $n\to\infty$ is the number of channel uses that will be used for transmitting $W$ from the EHU to the ET, and  $R$ denotes the data rate of   message $W$. We assume a priori knowledge of the CSI, i.e., $H_i$ is known for $i=1....n$ before the start of the communication at both nodes. Then,  we have
\begin{align}\label{eq_1}
nR& \leq H(W|H^n) = H(W|H^n)-H(W|H^n, Y_2^n,X_2^n)+H(W|H^n, Y_2^n,X_2^n)\nonumber\\
&=I(W;Y_2^n,X_2^n|H^n)+H(W|H^n, Y_2^n,X_2^n),
\end{align}
which follows from the definition of the mutual information. By Fano's inequality we have
\begin{align}\label{eq_2}
H(W|H^n, Y_2^n,X_2^n) \stackrel{(a)}{\leq} H(W|Y_2^n,H^n) \leq P_e n R+1,
\end{align}
where $(a)$ follows from the fact that conditioning reduces entropy and $P_e$ is the average probability of error of the message $W$.
Inserting (\ref{eq_2}) into (\ref{eq_1}), and dividing both sides by $n$, we have
\begin{align}\label{eq_3}
R\leq \frac{1}{n}I(W; Y_2^n,X_2^n|H^n)+P_e  R+1/n.
\end{align}
Assuming that $n\to\infty$, and assuming that  $P_e\to 0$ as $n\to \infty$, (\ref{eq_3}) can be written as
\begin{align}\label{eq_4v}
R\leq \frac{1}{n} I(W; Y_2^n,X_2^n|H^n).
\end{align}
We represent the right hand side of (\ref{eq_4v}) as
\begin{align}\label{eq_5}
I(W;Y_2^n,X_2^n|H^n)&=I(W;Y_2^n|X_2^n,H^n)+I(W;X_2^n|H^n).
\end{align}
Now, since the transmit message $W$ is uniformly drawn from the message set $\{1,\,2,\,...,2^{nR}\}$ at the EHU, and the ET does not know which message the EHU transmits, the following holds
\begin{align}\label{eq_5a}
I(W;X_2^n|H^n)=0.
\end{align}
Inserting (\ref{eq_5a}) into (\ref{eq_5}), we have
\begin{align}\label{eq_6}
I(W;Y_2^n,X_2^n|H^n)&=I(W;Y_2^n|X_2^n,H^n)\nonumber\\
&\stackrel{(a)}{\leq}\sum_{i=1}^n I(W;Y_{2i}|Y_2^{i-1},X_2^n,H^n)\nonumber\\
&\stackrel{(b)}{=} \sum_{i=1}^n H(Y_{2i}|Y_2^{i-1},X_2^n,H^n)-\sum_{i=1}^n H(Y_{2i}|Y_2^{i-1},X_2^n,H^n, W)\nonumber\\
&\stackrel{(c)}{\leq} \sum_{i=1}^n H(Y_{2i}|X_2^n,H^n)-\sum_{i=1}^n H(Y_{2i}|Y_2^{i-1},X_2^n,H^n,W)\nonumber\\
&\stackrel{(d)}{\leq} \sum_{i=1}^n H(Y_{2i}|X_2^n,H^n)-\sum_{i=1}^n H(Y_{2i}|Y_2^{i-1},X_{1i},X_2^n,H^n, W),
\end{align}
where $(a)$ follows from the fact that the entropy of a collection of random
variables is less than the sum of their individual entropies [\ref{lit20}], $(b)$ is a consequence of the chain rule, and $(c)$ and $(d)$ follow from the fact that conditioning reduces entropy. Now, due to the memoryless assumption, it easy to see that $Y_{2i}$ is independent of all elements in the vectors $X_2^n$ and $H^n$ except the elements  $X_{2i}$ and $H_i$. Thereby, the following holds
\begin{align}\label{eq_6b}
H(Y_{2i}|X_2^n,H^n)=H(Y_{2i}|X_{2i},H_i).
\end{align}
Moreover, it is easy to see that given $X_{1i}$ and $X_{2i}$, $Y_{2i}$ is independent of $Y_2^{i-1}$, of all elements in the vector $X_2^n$ except the element  $X_{2i}$, of all the elements of the vector $H^n$ except $H_i$, and of $W$.
Thereby, the following holds
\begin{align}\label{eq_6a}
H(Y_{2i}|Y_2^{i-1},X_{1i},X_2^n,H^n,W)=H(Y_{2i}|X_{1i},X_{2i},H_i).
\end{align}
Inserting (\ref{eq_6b}) and (\ref{eq_6a}) into (\ref{eq_6}), we obtain
\begin{align}\label{eq_6c}
I(W;Y_2^n,X_2^n|H^n)
&\leq \sum_{i=1}^n H(Y_{2i}|X_{2i},H_i)-\sum_{i=1}^n H(Y_{2i}|X_{1i},X_{2i},H_i)\nonumber\\
&=\sum_{i=1}^n I(X_{1i};Y_{2i}|X_{2i},H_i).
\end{align}
Now, inserting (\ref{eq_6c}) into (\ref{eq_4v}), we have
\begin{align}\label{eq_7}
R\leq \frac{1}{n} \sum_{i=1}^n I(X_{1i};Y_{2i}|X_{2i},H_i) =I(X_{1};Y_{2}|X_{2},H).
\end{align}
Hence, an upper bound on the capacity is given by (\ref{eq_7}) when no additional constraints on $X_1$ and $X_2$ exist. However, in our case, we have two constraints on $X_1$ and $X_2$. One constraint is that $\mathcal{E}\{X_2^2\}\leq P_{ET}$, expressed by C1 in (\ref{cap_eq1}). The other constraint is given by (\ref{eq_4b}), expressed by C2 in (\ref{cap_eq1}), which limits the average power of the EHU to be less than the maximum average harvested power. Constraints C3 and C4 in (\ref{cap_eq1}) come from the definitions of probability distributions. Hence, by inserting C1, C2, C3, and C4 from (\ref{cap_eq1}) into (\ref{eq_7}) and maximizing with respect to $p(x_1, x_2|h)=p(x_1|x_2, h)p(x_2|h)$, we obtain that the capacity is upper bounded by (\ref{cap_eq1}). This proves the converse. In Subsection~\ref{Ach}, we prove that this upper bound can be achieved. Thus, the capacity of the considered channel is given by (\ref{cap_eq1}).

\subsection{Achievability of the Channel Capacity}\label{Ach}
The capacity achieving coding scheme for this channel is similar to the coding scheme for the AWGN fading channel with EH given in [\ref{intro_lit6}]. The proposed scheme is outlined in the following.

 The EHU wants to transmit message $W$ to the ET using the harvested  energy  from the ET. Message $W$ is assumed to be drawn uniformly at random from a message set $\{1,2...2^{nR}\}$. Thereby, message $W$ carries $nR$ bits of information, where $n\rightarrow\infty$.

In the following, we describe a method for transferring $n R$ bits of information from the EHU to the ET in $n+b$ channel uses,  where $R=C-\epsilon$, and   $\epsilon\to 0$ and     $n/(n+b)\to 1$ as $n\to\infty$. As a result, the information  from the EHU to the ET is transferred at rate $R=C$, as $n\to\infty$.

For the proposed achievability scheme, we assume that   the transmission is carried out in   $N+B$ time slots, where $N/(N+B)\to 1$ as $N\to\infty$. In each time slot, we use the channel $k$ times, where $k\to\infty$. The numbers  $N$, $B$, and $k$ are chosen such that $n=Nk$ and $b=Bk$ hold.  Moreover, we assume that message $W$ is represented in a binary form as a sequence of bits  that is stored at the EHU.

\emph{Transmissions at the ET:} In each time slot, the ET transmits the same symbol $x_2$ during the $k$ channel uses of the considered time slot. The value of the symbol $x_2$ depends only on the fading gain of the channel $h$ during the corresponding time slot, and it can be found from Theorem 2.

\emph{Receptions and transmissions at the EHU:} During the first few time slots, the EHU is silent and only harvests energy from the ET. The EHU will transmit for the first time only when it has harvested enough energy both for processing and transmission, i.e., only when its harvested energy accumulates to a level which is higher than $P_p+P_{\rm EHU}(x_2,h)$, where $h$ is the fading gain in the time slot considered for transmission. In that case, the EHU extracts $k R(h)$ bits from its storage, maps them to a Gaussian codeword with rate $R(h)$ and transmits that codeword to the ET. The rate of the codeword $R(h)$ is given by
\begin{align}\label{eq_n1}
R(h)=\frac{1}{2} \log \left(1+\frac{h^2 P_{EHU}\left(\sqrt{P_{ET}},h\right)}{\sigma_2^2 + P_{ET} \alpha_2}\right) - \epsilon
\end{align}
if (\ref{cap_eq2a}) holds. Otherwise, if (\ref{cap_eq2a}) does not hold,  rate of the codeword $R(h)$ is given by
\begin{align}\label{eq_n2}
R(h)=\frac{1}{2} \log \left(1+\frac{h^2 P_{EHU}\left(x_0(h),h\right)}{\sigma_2^2 + x_0^2(h) \alpha_2}\right) - \epsilon.
\end{align}

\emph{Receptions at the ET:} The ET is able to decode the transmitted codeword from the EHU since it is received via an AWGN channel with total AWGN variance of $\sigma_2^2 + P_{ET} \alpha_2$ and $\sigma_2^2 + x_0^2(h) \alpha_2$ for the rates in (\ref{eq_n1}) and (\ref{eq_n2}), respectively.

The EHU and the ET repeat the above procedure for $N+B$ time slots.

Let $\mathcal{N}$ denote a set comprised of the time slots during which the EHU has enough energy harvested and thereby transmits and let $\mathcal{B}$ denote a set comprised of the time slots during which the EHU does not have enough energy harvested and thereby it is silent. Let $N=|\mathcal{N}|$ and  $B=\mathcal{B}$, where $||$ denotes the cardinality of a set. Moreover, let  $h(i)$ denote the outcome of $H$ in the $i$-th time slot. Using the above notations, the rate achieved during the $N+B$ time slots is given by
\begin{align}\label{eqn3}
R=\lim_{(N+B)\to\infty}\frac{1}{N+B}\sum_{i\in\mathcal{N} }  R(h(i))=\lim_{(N+B)\to\infty} \frac{N}{N+B}\sum_{h\in\mathcal{H}} R(h) p(h).
\end{align}
Now, it is proven in [\ref{lit19}] that when the EHU is equipped with an unlimited battery capacity and when (\ref{eq_4b}) holds, $N/(N+B)\to 1$ as $(N+B)\to\infty$. Using this, (\ref{eqn3}) simplifies to
\begin{align}\label{eqn4}
R=\sum_{h\in\mathcal{H}} R(h) p(h),
\end{align}
which is the channel capacity.

\section{Numerical Results}
In this section, we illustrate examples of the capacity of the system model outlined above, and compare it with the achievable rates of a chosen benchmark scheme. In the following, we outline the system parameters, than we introduce the benchmark scheme, and finally we provide the numerical results.
\subsection{System Parameters}
We use the standard path loss model given by
\begin{align}\label{num_1}
\mathcal {E} \{H^2\}=\Omega_H=\left(\frac{c}{f_c 4 \pi}\right)^2 d^{-\gamma}
\end{align}
in order to compute the average power of the channel fading gains of the ET-EHU/EHU-ET link, where $c$ denotes the speed of light, $f_c$ is the carrier frequency, $d$ is the length of the link, and $\gamma$ is the path loss exponent. We assume that $\gamma=3$. The carrier frequency is equal to $2.4$ GHz, a value used in practice for sensor networks, and $d=10$ m or $d=20$ m. We assume a bandwidth of $B=100$ kHz and noise power of $-160$ dBm per hertz, which for $100$ kHz adds-up to a total noise power of $10^{-14}$ Watts. The energy harvesting efficiency coefficient $\eta$ is assumed to be equal to $0.8$. The system parameters are summarized in Table I. Throughout this section, we assume Rayleigh fading with average power $\Omega_H$ given by (\ref{num_1}).
\begin{table}[]
\centering
\caption{Simulation parameters}
\label{my-label}
\begin{tabular}{ll}
\hline
Parameter                  & Value                      \\ \hline
Speed of light $c$	       & 299 792 458 m / s          \\ \hline
Carrier frequency $f_c$	   & 2.4 GHz					\\ \hline
Bandwidth $B$              & 100 kHz                     \\ \hline
Noise power $\sigma^2$     & -160 dBm per Hz            \\ \hline
EH efficiency $\eta$       & 0.8                        \\ \hline
Path loss exponent $\gamma$& 3                          \\ \hline
Distance $d$       		   & 10 m $\vee$ 20 m												\\ \hline
Processing cost $P_p$      & -10 dBm $\sim$ 10 dBm                    \\ \hline
ET transmit power $P_{ET}$ & 0 dBm $\sim$ 35 dBm                 \\ \hline
\end{tabular}
\end{table}
\vspace{7mm}
\subsection{Benchmark Schemes}
For the benchmark scheme, we divide the transmission time into slots of length $T$. We assume that the EHU is silent and only harvests energy during a portion of the time slot, denoted by $t$. In the remainder of the time slot, $(T-t)$, the EHU only transmits information to the ET and does not harvest energy. Similarly, the ET transmits energy during $t$, but remains silent and receives information during $T-t$. In other words, both the EHU and the ET work in a HD mode. In this mode, the nodes are not impaired by self-interference, thus the harvested energy in channel use $i$ is given by
\begin{align}\label{bench_1}
%E_i=\text{max} \{ 0, \eta (H_{i}X_{2i} + \bar g_{1} X_{1i}+G_{1i} X_{1i})^2-P_p \},
E_i= \eta (H_{i}X_{2i})^2.
\end{align}
Again, we assume CSI knowledge at the ET and at the EHU, and in addition the EHU is also equipped with a battery with an unlimited storage capacity. Therefore, as per [\ref{lit19}], the EHU can also choose any amount of power for information transmission as long as its average extracted energy from the battery is smaller than $\mathcal{E}\{E_i\}$, where $E_i$ is given by (\ref{bench_1}). Considering the HD nature of the channel, the EHU can achieve its maximum rate given by

\begin{align}\label{num_2}
R_{HD}= \underset{t} {\text{max}} \,\,\, t \log \left(1+\frac {h^2 P_{EHU} (h)}{\sigma_2^2}\right),
\end{align}
where
\begin{align}
P_{EHU} (h) =\text{max} \Big\{0, \frac{1}{\lambda} - \frac{\sigma^2}{h^2} \Big\},
\end{align}
where $\lambda$ is found such that
\begin{align}
t\left(P_p + \left[\frac{e^{-\frac{\lambda \sigma^2}{\Omega_H}}}{\lambda}-\frac{\sigma^2 E_1\left(\frac{\lambda \sigma^2}{\Omega_H}\right)}{\Omega_H}\right]\right)=(T-t) \eta P_{ET} \Omega_H
\end{align}
holds.

\vspace{7mm}
\subsection{Numerical Examples}
\begin{figure}[tbp]
\centering
\includegraphics[width=5.7in,scale=4]{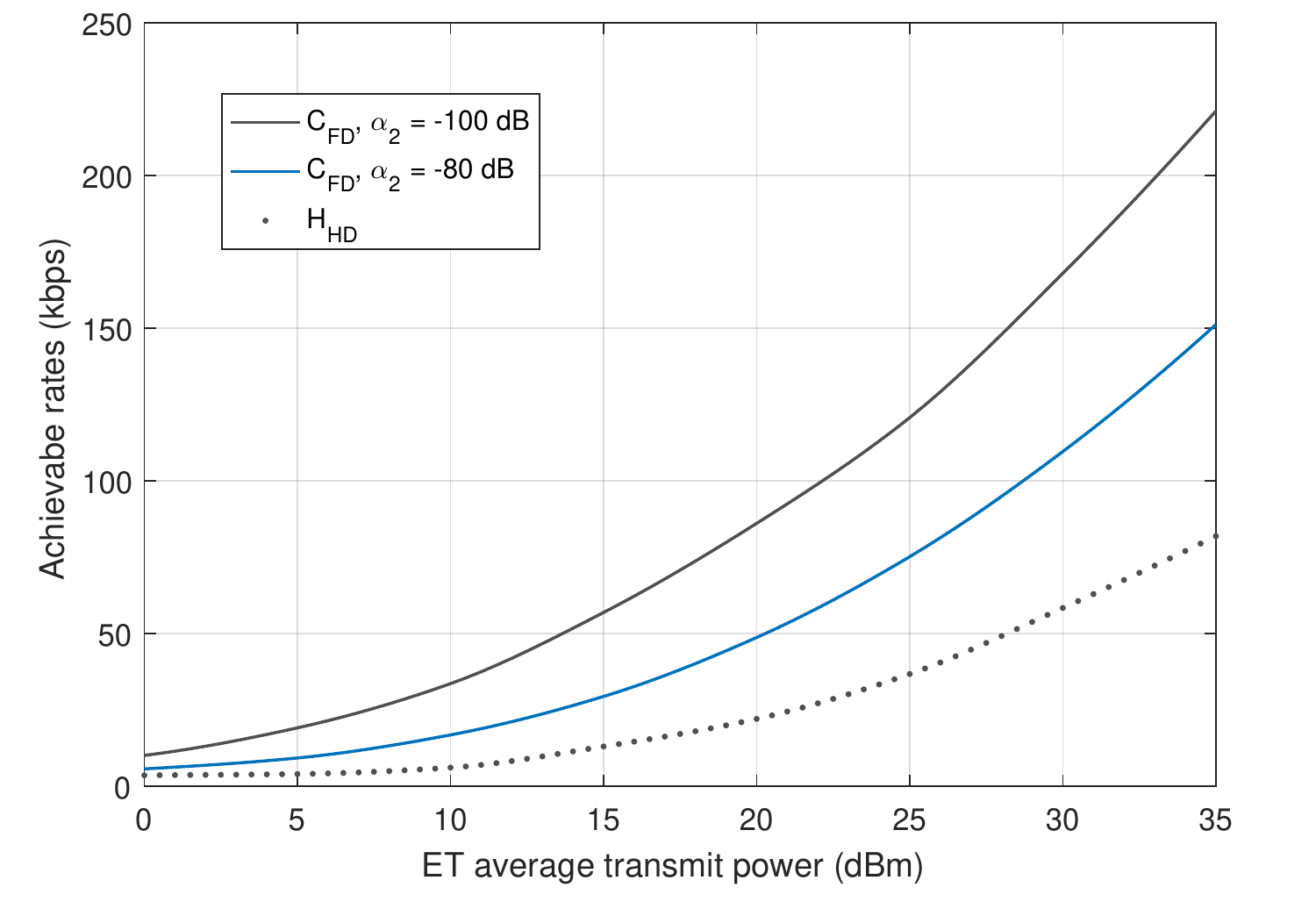}
\caption{Comparison of the capacity and the achievable rates of the benchmark scheme as a function of the ET average transmit power for a link distance of $d=10$m.}
\label{num1}
\end{figure}

\begin{figure}[tbp]
\centering
\includegraphics[width=5.7in,scale=4]{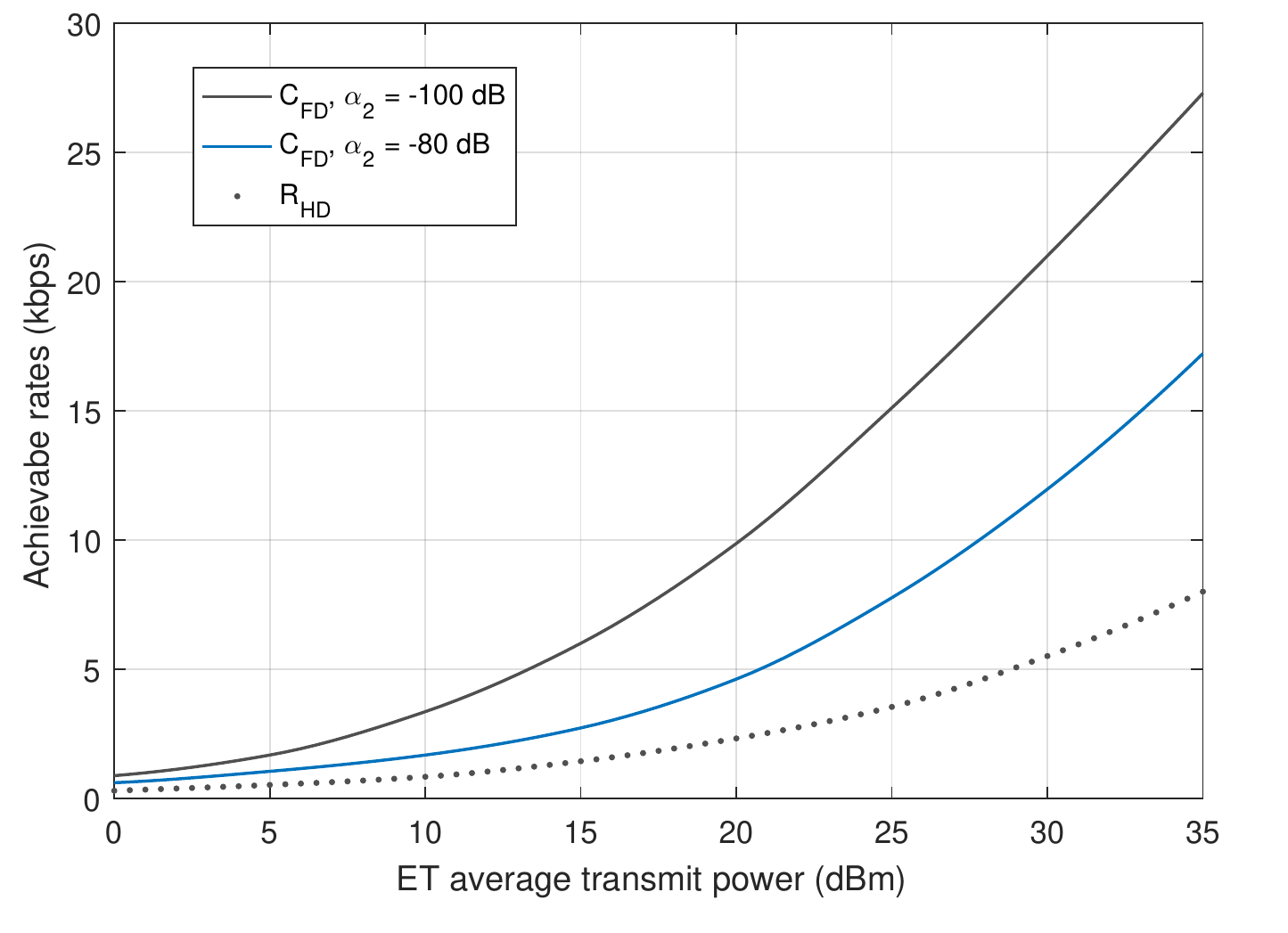}
\caption{Comparison of the capacity and the achievable rates of the benchmark scheme as a function of the ET average transmit power for a link distance of $d=20$m.}
\label{num2}
\end{figure}

Figs.~\ref{num1} and~\ref{num2} illustrate the data rates achieved with the proposed capacity achieving scheme and the benchmark scheme for link distances of $10$ m and $20$ m, respectively, and average power of the ET that ranges from $0$ dBm to $35$ dBm. The processing cost at the EHU is set to $P_p = -10$ dBm. It can be clearly seen from Figs. 3 and 4 that the achievable rates of the HD benchmark scheme are much lower than the derived channel capacity. The poor performance of the benchmark scheme is a consequence of the following facts. Firstly, self-interference energy recycling in the HD mode is impossible at the EHU since there is no self-interference. Secondly, the FD mode of operation is much more spectrally efficient than the HD mode, i.e., using part of the time purely for energy harvesting without transmitting information has a big impact on the system's performance. As it can be seen from comparing Figs.~\ref{num1} and~\ref{num2}, doubling the distance of the node, has a severe effect on performance.

\begin{figure}[tbp]
\centering
\includegraphics[width=5.7in,scale=4]{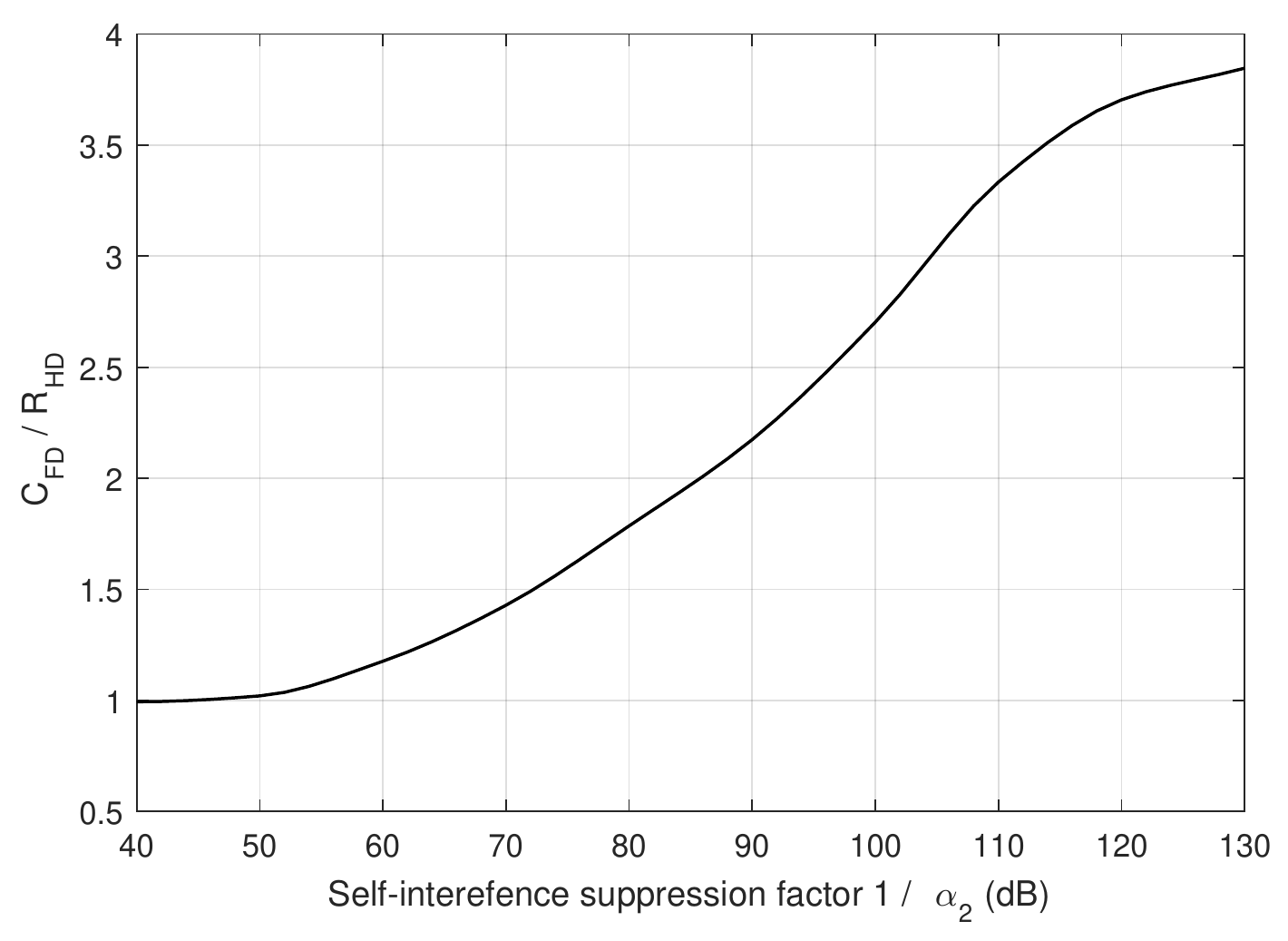}
\caption{$C_{FD}/R_{HD}$ ratio as a function of the self-interference suppression factor.}
\label{num3}
\end{figure}

In Fig.~\ref{num3}, we present the ratio between the capacity and the rate of the benchmark scheme, $C_{FD}/R_{HD}$, as a function of the self-interference suppression factor at the ET. The distance between the ET and the EHU is set to $d=10$ m and the average transmit power of the ET is set to $P_{ET}=30$ dBm. The self-interference suppression factor at the ET can be found as the reciprocial of the self-interference amplification factor $\alpha_2$, i.e., as $1/\alpha_2$. The ratio $C_{FD}/R_{HD}$ can be interpreted as the gain in terms of data rate obtained by using the proposed capacity achieving scheme compared to the data rate obtained by using the benchmark scheme. When the self-interference suppression factor is very small, i.e., around $40$ dB, it cripples the FD capacity and the FD rate converges to the HD rate. Naturally, as the self-interference is more efficiently suppressed, i.e., $\geq 50$ dB the FD capacity rate becomes significantly larger than the HD rate. An interesting observation can be made for suppression factor around $70$ dB, which are available in practice. In this case, the capacity achieving scheme results in a rate which is approximately $50\%$ larger than the HD rate. Another interesting result is that for self-interference suppression of more than $85$ dB, which is also available in practice, the proposed FD system model achieves a rate which is more than double the rate of the HD system. This effect is a result of the energy recycling at the EHU.

\begin{figure}[tbp]
\centering
\includegraphics[width=5.7in,scale=4]{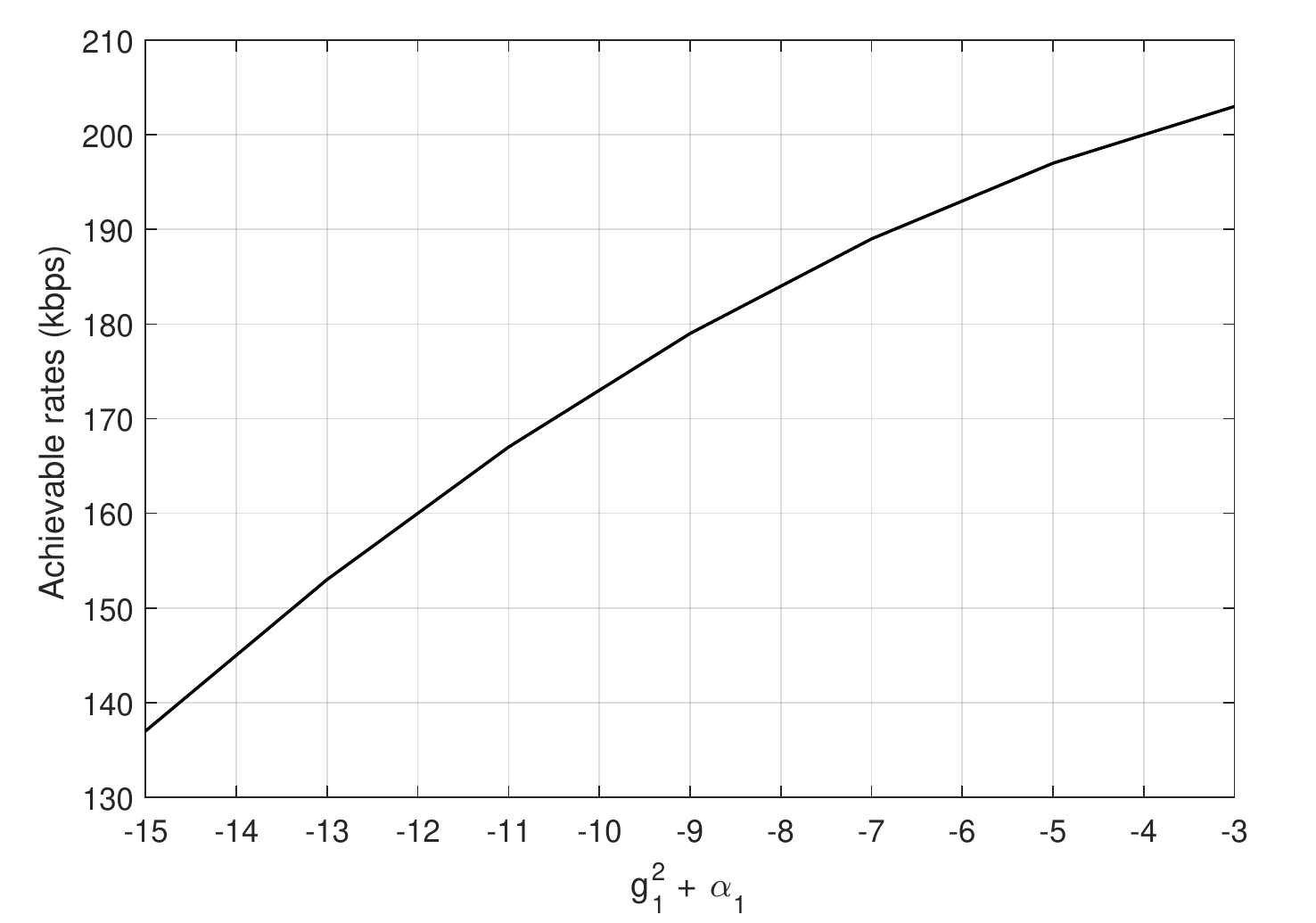}
\caption{Capacity as a function of $(\alpha_1+g_1^2)$.}
\label{num4}
\end{figure}

Fig.~\ref{num4} presents the capacity as a function of the mean and the variance of the self-interference channel at the EHU, $(\alpha_1+g_1^2)$. For this figure, the distance between the ET and the EHU is $d=10$ m and the average transmit power of the ET is $P_{ET}=30$ dBm. The self-interference suppression factor at the ET is $100$ dB. As the average self-interference channel gain at the EHU increases, i.e., $(\alpha_1+g_1^2)$ increases, the EHU can recycle a larger amount of its transmit energy. As a consequence, this results in a large capacity increase. Therefore, Figs.~\ref{num3} and~\ref{num4} illustrate that the self-interference at the EHU can be transformed from a deleterious factor, to an aide, or even an enabler of communication.

\begin{figure}[tbp]
\centering
\includegraphics[width=5.7in,scale=4]{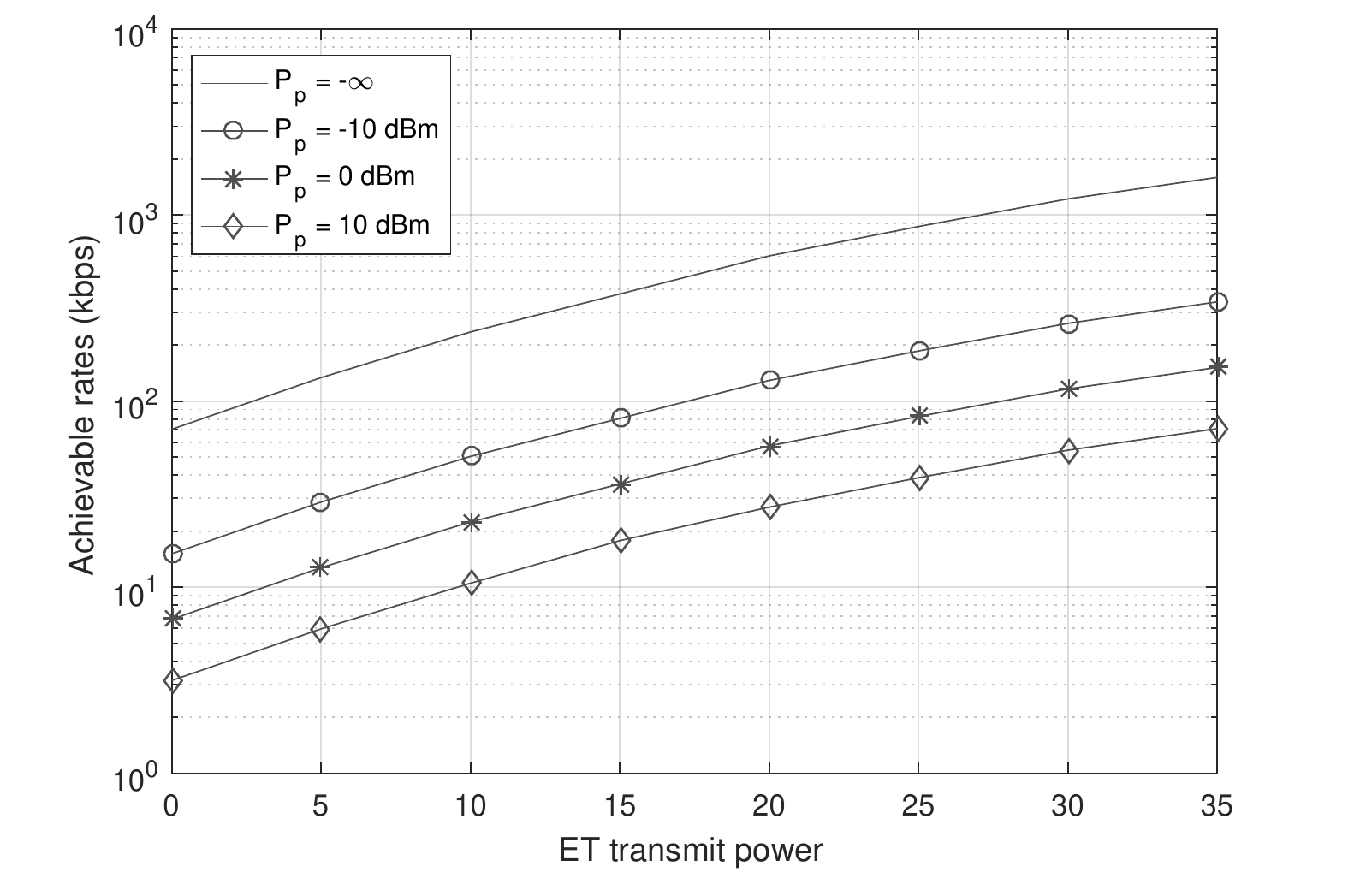}
\caption{Comparison of the capacity for different processing costs as a function of the ET's average transmit power for a link distance of $d=10$m.}
\label{num5}
\end{figure}

To illustrate the effect the processing energy cost has on the achievable rate, in Fig.~\ref{num5}, we present the capacity in the case when the processing cost is zero and non-zero, for a distance of $d=10$ m. The Y axes in Fig.~\ref{num5} is given in the logarithmic scale in order to better observe the discrepancy between the two scenarios. It can clearly be seen that when the processing cost is high, it has a detrimental effect on the capacity. This confirms that the energy processing cost must be considered in EH networks. Failing to do so might result in overestimating the achievable rates, which in reality would only represent very loose upper bounds that can never be achieved.

\section{Conclusion}
We studied the capacity of the point-to-point FD wirelessly powered communication system, comprised of an ET and an EHU. Because of the FD mode of operation, both the EHU and the ET experience self-interference, which impairs the decoding of the information-carrying signal at the ET, but serves as an additional energy source at the EHU. We showed that the capacity is achieved with a relatively simple scheme, where the input probability distribution at the EHU is zero-mean Gaussian and where the ET transmits only one symbol. Numerical results showed huge gains in terms of data rate when the proposed capacity achieving scheme is employed compared to a HD benchmark scheme as well as the indisputable effect that the processing cost and the self-interference have on the performance of WPCNs.

\section*{Acknowledgement}
The authors would like to thank Prof. P. Popovski, Aalborg University, Denmark, for comments that greatly improved the manuscript.

\appendices
\section{Proof of Theorem 2}
Let us assume that the optimal $p(x_2|h)$ is a discrete and that the optimal $p(x_1|x_2,h)$ is a continuous probability distribution, which will turn out to be valid assumptions. In order to find both input distributions, in the following, we solve the optimization problem given by (\ref{cap_eq1}).

Since $I(X_{1};Y_{2}|X_{2}=x_2, H=h)$ is the mutual information of an AWGN channel with channel gain $h$ and AWGN with variance $\sigma_2^2 + x_2^2 \alpha_2$, the optimal input distribution at the EHU, $p(x_1|x_2,h)$, is Gaussian with mean zero and variance $P_{EHU}(x_2,h)$, which has to satisfy constraint C2 in (\ref{cap_eq1}). Thereby, $I(X_{1};Y_{2}|X_{2}=x_2, H=h)=\frac {1}{2} \log \left(1+\frac{h^2 P_{EHU}\left(x_2,h\right)}{\sigma_2^2 + x_2^2 \alpha_2}\right)$. Now, since $G_1$ and $X_1$ are zero-mean Gaussian RVs, the left-hand side of constraint C2 can be transformed into
\begin{flalign} \label{appB_eq0}
&\int_{x_1} \sum_{x_2 \in \mathcal{X}_2} \sum_{h \in \mathcal{H}} (x_1^2 + P_p) p(x_1|x_2,h)p(x_2|h)p(h)dx_1  \\ \notag
&=\sum_{x_2 \in \mathcal{X}_2} \sum_{h \in \mathcal{H}} P_{EHU} (x_2,h) p(x_2|h)p(h) +P_p.
\end{flalign}
Whereas, the right-hand side of C2 can be rewritten as
\begin{flalign}\label{appB_eq0aa}
&\int_{x_1} \sum_{x_2 \in \mathcal{X}_2} \sum_{h \in \mathcal{H}} E_{\rm in} p(x_1|x_2,h)p(x_2|h)p(h)d x_1  \\ \notag
&=\int_{g_1} \int_{x_1} \sum_{x_2 \in \mathcal{X}_2} \sum_{h \in \mathcal{H}} \eta(h x_2 + \bar g_{1} x_{1}+g_{1} x_{1})^2 p(x_1|x_2,h)p(x_2|h)p(h)p(g_1)d x_1 d g_1  \\ \notag
&=\sum_{x_2 \in \mathcal{X}_2} \sum_{h \in \mathcal{H}} \eta h^2 x_{2}^2 p(x_2|h)p(h) + \int_{x_1}\sum_{x_2 \in \mathcal{X}_2} \sum_{h \in \mathcal{H}} \eta \bar g_{1}^2 x_{1}^2 p(x_1|x_2,h)p(x_2|h)p(h)d x_1 \\ \notag
&+\int_{g_1} \int_{x_1}\sum_{x_2 \in \mathcal{X}_2} \sum_{h \in \mathcal{H}} \eta g_{1}^2 x_{1}^2 p(x_1|x_2,h)p(x_2|h)p(h)p(g_1)d x_1 d g_1 = \sum_{x_2 \in \mathcal{X}_2} \sum_{h \in \mathcal{H}} \eta h^2 x_{2}^2 p(x_2|h)p(h)  \\ \notag
&+\eta \bar g_{1}^2 \sum_{x_2 \in \mathcal{X}_2} \sum_{h \in \mathcal{H}} P_{EHU} (x_2,h) p(x_2|h)p(h) + \eta \alpha_1 \sum_{x_2 \in \mathcal{X}_2} \sum_{h \in \mathcal{H}} P_{EHU} (x_2,h) p(x_2|h)p(h),
\end{flalign}
where $g_1$ represents the realizations of the random variable $G_1$. Combining (\ref{appB_eq0}) and (\ref{appB_eq0aa}) transforms (\ref{cap_eq1}) into
\begin{align} \label{appB_eq0a}
&\max_{P_{EHU}(x_2,h), p(x_2|h)} \sum_{x_2 \in \mathcal{X_2}} \sum_{h \in \mathcal{H}}{ \frac{1}{2} \log \left(1+\frac{h^2 P_{EHU}(x_2,h)}{\sigma_2^2 + x_2^2 \alpha_2} \right) p(x_2|h)p(h)}\nonumber\\
&{\rm{Subject\;\;  to}}  \nonumber\\
&\qquad\qquad{\rm C1:} \sum_{x_2 \in \mathcal{X}_2} \sum_{h \in \mathcal{H}} x_2^2 p(x_2|h)p(h) \leq P_{ET} \nonumber\\
&\qquad\qquad{\rm C2:} \sum_{x_2 \in \mathcal{X}_2} \sum_{h \in \mathcal{H}} P_{EHU}(x_2,h)p(x_2|h)p(h) + P_p \leq \nonumber\\
&\qquad\qquad\qquad  \sum_{x_2 \in \mathcal{X}_2} \sum_{h \in \mathcal{H}} \eta h^2 x_2^2 p(x_2|h)p(h) + \eta(\bar {g_1}^2 +\alpha_1)\sum_{x_2 \in \mathcal{X}_2} \sum_{h \in \mathcal{H}} P_{EHU}(x_2,h)p(x_2|h)p(h) \nonumber\\
&\qquad\qquad{\rm C3:} \sum_{x_2 \in \mathcal{X}_2}{p(x_2|h)} = 1 \nonumber\\.
&\qquad\qquad{\rm C4:} P_{EHU}(x_2,h)\geq 0 .\nonumber\\
\end{align}
Now, (\ref{appB_eq0a}) can be solved in a straightforward manner using the Lagrange duality method. Thereby, we write the Lagrangian of (\ref{appB_eq0a}) as
\begin{flalign} \label{appB_eq0b}
\mathcal{L}&=\sum_{x_2 \in \mathcal{X_2}}\sum_{h \in \mathcal{H}}{\frac{1}{2} \log \left(1+\frac{h^2 P_{EHU}(x_2,h)}{\sigma_2^2 + x_2^2 \alpha_2} \right) p(x_2|h)p(h)}  \\ \notag &-\lambda_1 \left(\sum_{x_2 \in \mathcal{X_2}} \sum_{h \in \mathcal{H}} {x_2^2 p(x_2|h)p(h)}-P_{ET}\right) -\mu_1 \left(\sum_{x_2 \in \mathcal{X_2}}{p(x_2|h)}-1\right) - \mu_2 P_{EHU} (x_2,h)  \\ \notag
&-\lambda_2 \left((1-\eta(\bar {g_1}^2 +\alpha_1))\sum_{x_2 \in \mathcal{X}_2} \sum_{h \in \mathcal{H}}P_{EHU}(x_2,h)p(x_2|h)p(h) + P_p -\sum_{x_2 \in \mathcal{X}_2} \sum_{h \in \mathcal{H}} \eta h^2 x_2^2 p(x_2|h)p(h)\right). \\ \notag
\end{flalign}
In (\ref{appB_eq0a}), we assume that $0<\eta(\bar {g_1}^2 +\alpha_1)<1$, since $\eta(\bar {g_1}^2 +\alpha_1)\geq1$ would practically imply that the EHU recycles the same or even a larger amount of energy than what has been transmitted by the EHU, which is not possible in reality. In (\ref{appB_eq0b}), $\lambda_1$, $\lambda_2$, $\mu_1$, and $\mu_2$ are the Lagrangian multipliers associated with C1, C2, C3, and C4 in (\ref{cap_eq1}), respectively. Differentiating (\ref{appB_eq0b}) with respect to the optimization variables, we obtain
\begin{align}
\frac{\partial \mathcal{L}}{\partial P_{EHU}(x_2,h)}&=\frac{\frac{h^2}{\sigma_2^2 + x_2^2 \alpha_2}}{1+\frac{h^2 P_{EHU}(x_2,h)}{\sigma_2^2 + x_2^2 \alpha_2}} -\lambda_2 (1-\eta(\bar {g_1}^2 +\alpha_1)) -\mu_2=0 \label{appB_eq0c} \\
\frac{\partial L}{\partial p(x_2|h)}&=\frac{1}{2} \sum_{h \in \mathcal{H}} \log \left(1+\frac{h^2 P_{EHU}(x_2,h)}{\sigma_2^2 + x_2^2 \alpha_2}\right) p(h) - \lambda_1 \sum_{h \in \mathcal{H}}x_2^2p(h) - \mu_1  \nonumber\\
&-\lambda_2 \left((1-\eta(\bar {g_1}^2 +\alpha_1))\sum_{h \in \mathcal{H}}P_{EHU}(x_2,h)p(h)- \eta \sum_{h \in \mathcal{H}}h^2 x_2^2 p(h)\right)=0. \label{appB_eq2}
\end{align}
When $P_{EHU}(x_2,h)>0$, then $\mu_2=0$. In consequence, we can use (\ref{appB_eq0c}) to find $P_{EHU}(x_2,h)$ as given by Theorem 2.

When $\mu_1 > 0$, then (\ref{appB_eq2}) has only two possible solutions $\pm x_2^*$. In order for $p^*(x_2|h)$ to be a valid probability distribution $p(x_2^*|h)+p(-x_2^*|h)=1$ has to hold. Furthermore, $\sum_{x_2 \in \mathcal{X}_2} {x_2^2 p(x_2|h) p(h)} \leq P_{ET}$ also has to hold. So, one possible solution for the optimal input distribution is $p^*(x_2|h)=\frac{1}{2} \delta(x_2-x_2^*)+\frac{1}{2} \delta(x_2+x_2^*)$. However, $x_2$ does not need to carry any information to the EHU, thus uniformly choosing $x_2$ between $x_2^*$ and $-x_2^*$ brings no benefit to the EHU. Moreover, the complexity of the ET will be reduced if it only transmits one symbol, the symbol $x_2^*$. Thus, $p^*(x_2|h)=\delta(x_2-x_2^*)$ is chosen. Now for $\sum_{x_2 \in \mathcal{X}_2} {x_2^2 p(x_2|h) p(h)} \leq P_{ET}$ to hold, we can choose $x_2^*=\sqrt{P_{ET}}$. Thus possible solution for $p^*(x_2|h)$ is given by (\ref{cap_eq22a}) in Theorem 2. Using (\ref{appB_eq2}), we find the condition for $p(x_2|h)=\delta(x_2-\sqrt{P_{ET}})$ as
\begin{align}\label{appB_eq3}
&\sum_{h \in \mathcal{H}} \frac{1}{2} \log \left(1+\frac{h^2 P_{EHU}(\sqrt{P_{ET}},h)}{\sigma_2^2 + P_{ET} \alpha_2}\right)p(h) = \lambda_1 P_{ET} + \mu_1  \\ \notag
&+\lambda_2 \left(\left(1-\eta(\bar {g_1}^2 +\alpha_1)\right)\sum_{h \in \mathcal{H}}P_{EHU} \left(\sqrt{P_{ET}},h\right)p(h)- \eta P_{ET} \sum_{h \in \mathcal{H}}h^2 p(h)\right).
\end{align}
In this case, the capacity is given by (\ref{cap_eq3}).
On the other hand, when (\ref{appB_eq3}) does not hold, another possible way to satisfy $\sum_{x_2 \in \mathcal{X}_2} {x_2^2 p(x_2|h) p(h)} \leq P_{ET}$ is to have $x_2^*=x_0(h)$, where in addition to C1 in (\ref{appB_eq0a}), the following has to be satisfied
\begin{align}\label{appB_eq4}
&\sum_{h \in \mathcal{H}} \frac{1}{2} \log \left(1+\frac{h^2 P_{EHU}(x_0(h),h)}{\sigma_2^2 + x_0^2(h) \alpha_2}\right)p(h) = \lambda_1 \sum_{h \in \mathcal{H}} x_0^2(h) p(h) + \mu_1 +\\ \notag
&\lambda_2 \left(\left(1-\eta(\bar {g_1}^2 +\alpha_1)\right) \sum_{h \in \mathcal{H}}P_{EHU}(x_0(h),h)p(h)- \eta\sum_{h \in \mathcal{H}} x_0^2(h) h^2  p(h) \right).
\end{align}
Using (\ref{appB_eq4}) and C3 in (\ref{appB_eq0a}), we can find the optimal $x_0(h)$ as given by (\ref{cap_eq2b}) in Theorem 2.

\end{document}